\documentclass[%
reprint,
superscriptaddress,
 amsmath,amssymb,
 aps,
 prl,
]{revtex4-1}

\usepackage{amsmath}    
\usepackage{graphicx}   
\usepackage{verbatim}   
\usepackage{color}      
\usepackage{subfigure}  
\usepackage{hyperref}   
\usepackage[normalem]{ulem} 
\newcommand{\ttilde}{\raise.17ex\hbox{$\scriptstyle\sim$}}
\definecolor{green}{rgb}{0.0, 0.42, 0.24}

\begin{document}

\title{Frustrated Coulomb explosion of small helium clusters}

\author{S. Kazandjian}
\affiliation{Sorbonne Universit\'{e}, CNRS, Laboratoire de Chimie Physique - Mati\`{e}re et Rayonnement, UMR 7614, F-75005 Paris, France}
\author{J. Rist}
\author{M. Weller}
\author{F. Wiegandt}
\author{D. Aslit\"urk}
\author{S. Grundmann}
\author{M. Kircher}
\author{G. Nalin}
\author{D. Pitters}
\author{I. Vela P\'erez}
\author{M. Waitz}
\affiliation{Institut f\"ur Kernphysik, J.~W.~Goethe Universit\"at, Max-von-Laue-Str. 1, D-60438 Frankfurt, Germany}
\author{G. Schiwietz}
\affiliation{Helmholtz-Zentrum Berlin f\"ur Materialien und Energie, Division NP-ABS, Hahn-Meitner-Platz 1, D-14109 Berlin, Germany}
\author{B. Griffin}
\author{J. B. Williams}
\affiliation{Department of Physics, University of Nevada Reno, 1664 N. Virginia Street Reno, NV 89557, USA}
\author{R. D\"orner}
\author{M. Sch\"offler}
\affiliation{Institut f\"ur Kernphysik, J.~W.~Goethe Universit\"at, Max-von-Laue-Str. 1, D-60438 Frankfurt, Germany}
\author{T. Miteva}
\affiliation{Sorbonne Universit\'{e}, CNRS, Laboratoire de Chimie Physique - Mati\`{e}re et Rayonnement, UMR 7614, F-75005 Paris, France}
\author{F. Trinter}
\affiliation{Institut f\"ur Kernphysik, J.~W.~Goethe Universit\"at, Max-von-Laue-Str. 1, D-60438 Frankfurt, Germany}
\affiliation{Deutsches Elektronen-Synchrotron (DESY), FS-PE, Notkestrasse 85, D-22607 Hamburg, Germany}
\affiliation{Fritz-Haber-Institut der Max-Planck-Gesellschaft, Molecular Physics, Faradayweg 4, 14195 Berlin, Germany}
\author{T. Jahnke}\email{jahnke@atom.uni-frankfurt.de}
\affiliation{Institut f\"ur Kernphysik, J.~W.~Goethe Universit\"at, Max-von-Laue-Str. 1, D-60438 Frankfurt, Germany}
\author{N. Sisourat}\email{nicolas.sisourat@upmc.fr}
\affiliation{Sorbonne Universit\'{e}, CNRS, Laboratoire de Chimie Physique - Mati\`{e}re et Rayonnement, UMR 7614, F-75005 Paris, France}

\date{\today}

\begin{abstract}
Almost ten years ago, energetic neutral hydrogen atoms were detected after a strong-field double ionization of H$_2$. This process, called 'frustrated tunneling ionization', occurs when an ionized electron is recaptured after being driven back to its parent ion by the electric field of a femtosecond laser. In the present study we demonstrate that a related process naturally occurs in clusters without the need of an external field: we observe a charge hopping that occurs during a Coulomb explosion of a small helium cluster, which leads to an energetic neutral helium atom. This claim is supported by theoretical evidence. As an analog to 'frustrated tunneling ionization', we term this process 'frustrated Coulomb explosion'.

\end{abstract}

\maketitle

Charging a molecule or cluster can lead to a Coulomb explosion as the charged constituents repel each other via the Coulomb force. This widespread phenomenon occurs, for example, as a consequence of stripping off electrons as molecular ion beams traverse a foil \cite{Vager_1989,Fadanelli_2004}, of multiple ionization induced by Auger cascades \cite{pitzer_absolute_2016,guillemin_selecting_2015}, of multiple ionization by charge transfer \cite{neumann_fragmentation_2010}, free electron lasers radiation \cite{erk_imaging_2014} or strong femtosecond laser pulses \cite{legare_laser_2006,pitzer_direct_2013}. Coulomb explosions have been successfully used to image static molecular \cite{weber_complete_2004,pitzer_direct_2013} or cluster structures \cite{zeller_imaging_2016, kunitski_m._observation_2015} and to follow structural changes \cite{boll_charge_2016} or electronic transitions \cite{trinter_evolution_2013} in real time.

A peculiarity concerning Coulomb explosions triggered by strong laser fields has been reported almost ten years ago. In a series of publications it has been demonstrated that in strong-field ionization processes, not only charged but also neutral energetic particles can be generated. A first publication by Eichmann \emph{et al.} reported on a corresponding observation in atoms \cite{Eichmann2009Nat} and a later work extended that concept to the molecular case and the Coulomb explosion process \cite{manschwetus_strong_2009}. In that work the occurrence of neutral hydrogen atoms with kinetic energy of several eV was noticed in a 'Coulomb explosion without double ionization'. In all cases the strong laser field is vital for the process: after  tunneling ionization the emitted electron is driven back by the laser field and recaptured by its parent ion. Even though the ion is neutralized, it still has the kinetic energy obtained from its previous acceleration in the laser field or from a Coulomb explosion that occurred in the molecular case. As a result, surprisingly energetic neutral particles are emitted and it became obvious that this process, which has been termed 'frustrated tunneling ionization', is actually a very common route in the interaction of strong laser fields with matter.

In the present letter we demonstrate that a related process may occur in loosely bound matter, as for examples clusters bound by the van der Waals forces, even without the need of an external strong laser field. As a charged particle emerges from the inner bulk of a cluster, it may collide with other atoms of the cluster and transfer parts of its kinetic energy to neutral fragments by elastic scattering. This route of energy transfer is routinely considered, especially in case of large clusters or droplets. We report here on another, very efficient, process: if the charged particle passes by another atom of the cluster, the ion can capture an electron from the atom and the charge is thus transferred between the two collision partners. In that case, a charged ion with almost no kinetic energy is generated, while the initially ionized particle is now neutral and possesses the kinetic energy it acquired during its time as an ion. Especially in case of a Coulomb explosion, where ions obtain large amounts of kinetic energy due to the Coulomb repulsion, energetic neutral fragments appear due to this charge hopping scenario. It should be mentioned that the charge hopping process has been already investigated in the case of singly-ionized helium droplets, leading to He$_2^+$ formation. However, in the latter case the charge hopping is a purely electronic process since it is faster than the nuclear motion \cite{halberstadt_resonant_1998,buchta_extreme_2013}. In the present case, nuclear motion is fast due to Coulomb explosion and the charge hopping does not lead to  He$_2^+$ but to a fast neutral atom and a slow ion.

In order to investigate a possible occurrence of this 'aborted' or 'frustrated' Coulomb explosion, a shake-up ionization of small helium clusters of size $m$ ($m < 5$) to the $(n=2)$-excited states and subsequent Interatomic Coulombic Decay (ICD) \cite{cederbaum_giant_1997,Marburger2003prl,Jahnke2004prl} is triggered employing synchrotron radiation of $h\nu = 66.4$~eV photon energy:

\begin{equation}
\mathrm{He}_m \xrightarrow{h\nu} \mathrm{He}_m^{+*}(n=2) + e_{\text{ph}}^-
\end{equation}

 As ICD occurs, electronic excitation energy is transferred from the excited atom to one of its loosely bound neighbors, which emits the received energy by releasing a second electron \cite{Hergenhahn11,Hergenhahn12,Jahnke15}. After the process, two positive charges are facing each other and the system fragments rapidly in a Coulomb explosion. Due to the weak van der Waals binding, the strong acceleration by the Coulomb explosion, and the overall small cluster size, the resulting ionic fragments typically consist of single atoms instead of larger fragments:

 \begin{equation}
\mathrm{He}_m^{+*}(n=2) \xrightarrow{\mathrm{ICD}} \mathrm{He}^{+} + \mathrm{He}^{+} + \mathrm{He}_{m-2} + e_{\text{ICD}}^-
\end{equation}

In case of helium dimers this ICD-route has been investigated in large detail by experiments and in theory \cite{havermeier_interatomic_2010,sisourat_ultralong-range_2010,trinter_evolution_2013,burzynski_interatomic-coulombic-decay-induced_2014,fruhling_time-resolved_2015}. From these studies the expected ion kinetics are well known. Because of the weak binding forces, it can be expected that the fragmentation into two singly charged He ions and a residual neutral cluster should lead in principle to very similar kinetics, which are dominated by the Coulomb explosion of the two He$^+$ ions. Accordingly, the combination of shake-up ionization and subsequent ICD is a suitable tool to introduce two charges in a cluster in a very well defined manner and investigate the dynamics of the cluster fragmentation.

In this combined experimental-theoretical study, we employed a Cold Target Recoil Ion Momentum Spectroscopy (COLTRIMS) setup \cite{dorner_cold_2000,ullrich_recoil-ion_2003,jahnke_multicoincidence_2004} in order to investigate the ejection of He$^+$ ions from small He clusters. The experimental setup was similar to the one described in \cite{havermeier_interatomic_2010}. Briefly, a supersonic gas jet (precooled to approx. 8~K) was intersected with photons ($h\nu = 66.4$~eV) from the synchrotron lightsource BESSY II in Berlin \cite{Schiwietz_2015}. Charged fragments (electrons and ions) created in a photoreaction were guided by weak electric and magnetic fields towards two time- and position-sensitive particle detectors \cite{jagutzki_multiple_2002}. By measuring the flight times and the positions of impact on the detectors, the particle trajectories inside the COLTRIMS spectrometer were reconstructed yielding the initial vector momenta of all charged particles. The ion arm of the spectrometer consisted of a short acceleration region of  3~cm length, while the electron arm incorporated a Wiley-McLaren time-focussing scheme \cite{wiley_timeflight_1955} consisting of an acceleration region (5~cm) with an electric field strength of 6~V/cm and a field-free drift region (10 cm in length). A superimposed homogeneous magnetic field (7~Gauss) yielded a full solid angle of detection of electrons of up to 15~eV kinetic energy. By measuring the momenta of all emitted electrons and ions in coincidence, reactions of clusters in which ICD occurred were discriminated from He monomer reactions. Furthermore, the emission angles in the laboratory frame of all particles are deduced from the measured momenta and from these relative emission angles between detected particles can be inferred, as well. By changing the He stagnation pressure and the gas nozzle temperature, the condensation properties of the supersonic expansion can be adjusted such, that apart from monomers (i.e. helium atoms that do not condensate at all) the condensed part of the gas jet is mainly a mixture of He$_2$ and He$_3$ or e.g. a mixture of slightly larger clusters (He$_3$, He$_4$).

\begin{figure}[h]
	\centering
	\includegraphics[width=\linewidth]{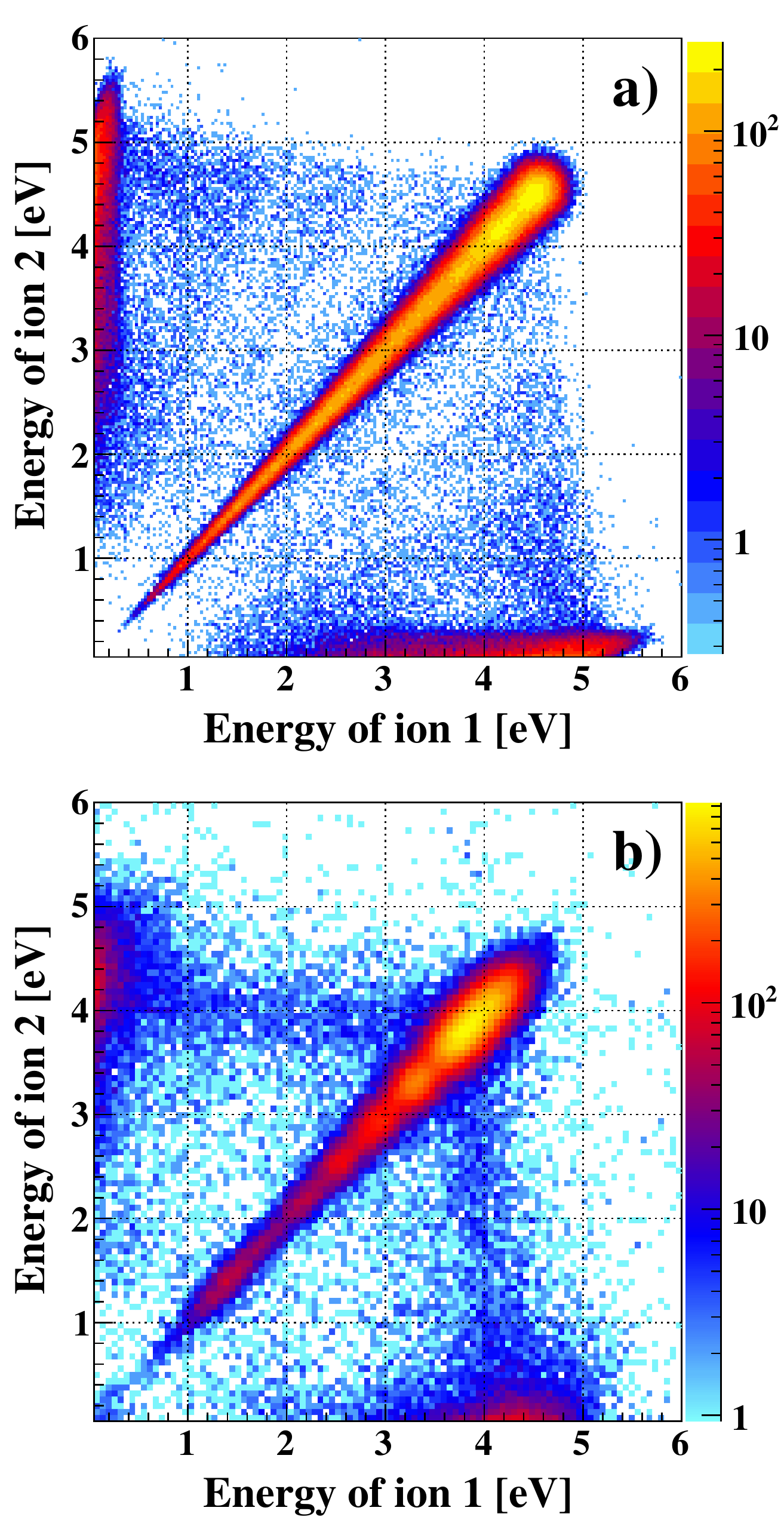}
	\caption{\label{fig:map} Coincidence map of ion kinetic energies occurring after the shake-up ionization and ICD. a) Theoretical results for He$_3$, b) measured distribution using a mixture of small clusters consisting mainly of He$_3$.}
\end{figure}

The ICD process and the subsequent Coulomb explosion were simulated with a semiclassical approach combined with a diatomics-in-molecules (DIM) technique. Both are described in detail in \cite{sisourat_nuclear_2013} and \cite{sisourat_interatomic_2016}, respectively. In brief, the motions of the nuclear quantum wave packets are replaced by a swarm of classical trajectories. Each trajectory propagates on the potential energy surface of one of the excited states until a decay condition is met \cite{sisourat_nuclear_2013}. After the decay, the trajectories are further propagated on the potential energy surface of one of the doubly ionized states. Each trajectory can hop from one surface to another according to a Landau-Zener probability (see \cite{Xie_jcp2017}). Furthermore, a Mulliken population analysis is performed on the DIM eigenvectors to obtain the charge of each atom along the trajectories. The initial conditions are obtained according to the $\mathrm{He}_m$ ground state nuclear wave functions by Rick \emph{et al.} \cite{rick_variational_1991}, and the starting potential energy surface is drawn uniformly among all the $\mathrm{He}^{+*}_m$ electronic excited states. The energy gradients and the ICD rates needed for the propagation are obtained from the DIM technique  \cite{sisourat_interatomic_2016}. The diatomic energies and the ICD rates of He$_2^{+*}(2p)$ states were taken from \cite{kolorenc_interatomic_2010}. He$_2^{+*}(2s)$ states were neglected since they contribute less to ICD \cite{sisourat_ultralong-range_2010}. For computing the final states, the energies of the lowest states of He$_2^{+}$ were taken from \cite{xie_accurate_2005} and we used a Coulombic potential for each pair of He$^+$-He$^+$ as these are valid for all ICD-relevant distances \cite{kolorenc_interatomic_2010}. The He-He fragment potential energy was taken from \cite{tang_accurate_1995}. The experiment has a constant momentum resolution, and hence an energy resolution which depends linearly on the measured energy. In order to emulate this energy resolution, our theoretical spectra have been convoluted with a Gaussian function with an energy-dependent width.

Figure~\ref{fig:map} depicts a coincidence map of ion kinetic energies occurring after the shake-up ionization and subsequent ICD. A good agreement between the simulation and the experiment is seen. The map can be divided into three parts: a narrow diagonal feature, weak and continuous vertical/horizontal lines for one ion kinetic energy around 4-5 eV, and two surprisingly strong islands where one of the ions has a kinetic energy close to zero. The diagonal line corresponds to a dimer-like Coulomb explosion where the two ions repel each other and do not exchange energy with other surrounding atoms. It turns out that the two other features are signatures of energy or charge transfer processes between one ion and a neutral helium atom within the cluster.

\begin{figure}
	\centering
	\includegraphics[scale=0.25,width=\linewidth]{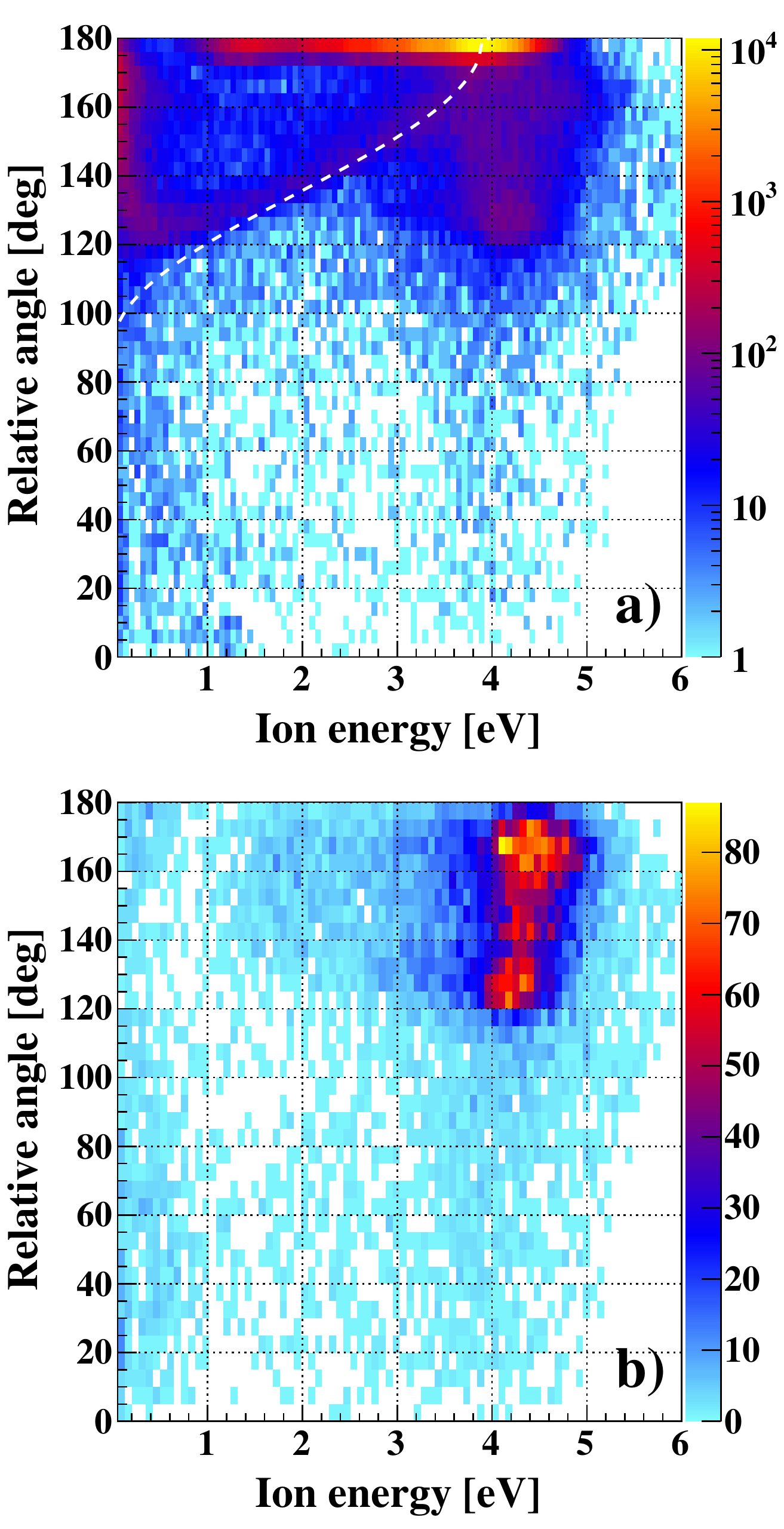}
	\caption{\label{fig:angle} Measured ion kinetic energy versus relative emission angle between both ions. a) Full dataset, b) events where one of the ions was detected with low kinetic energy (see text). The white, dashed line in a) shows the deflection-angle-dependent energy loss of a perfect elastic scattering event of an ion with an initial energy of 3.9~eV colliding with a neutral He atom initially at rest.}
\end{figure}

In order to further investigate the two non-diagonal features observed in Fig.~\ref{fig:map}, we show in Fig.~\ref{fig:angle} a) the energy of one of the ions versus the relative angle between both measured ions. Here we show the experimental data, which is well reproduced by the calculations (not shown). The most intense feature results from emission with a relative angle of 180$^\circ$ between the two ions. These events belong to an undisturbed (i.e. dimer-like) Coulomb explosion. Furthermore, a curved line occurs (highlighted by the white, dashed line): a classical binary collision leads to a distinct energy transfer, which is (for a fixed initial energy) solely dependent on the scattering angle after the collision. Such a dependency has been seen recently in \cite{Wiegandt2018arxiv} and such elastic scattering processes have been identified previously already by Shcherbinin \textit{et al}. They observed an energy loss appearing in the kinetic energy distributions of the ions generated after ICD in large helium clusters (between 1200 and 27000 atoms) and suggested an elastic collision mechanism (one ion transfers its kinetic energy to a neutral atom) in order to explain its occurrence \cite{shcherbinin_interatomic_2017}. Finally, there is an island at ion kinetic energy around 4-5 eV and emission angles between 120$^\circ$ and 180$^\circ$. This broad feature survives when restricting the dataset to events where one ion has a kinetic energy close to zero, as demonstrated by Fig.~\ref{fig:angle} b). These are events of frustrated Coulomb explosion as we demonstrate below.

\begin{figure}[h]
	\includegraphics[width=\linewidth]{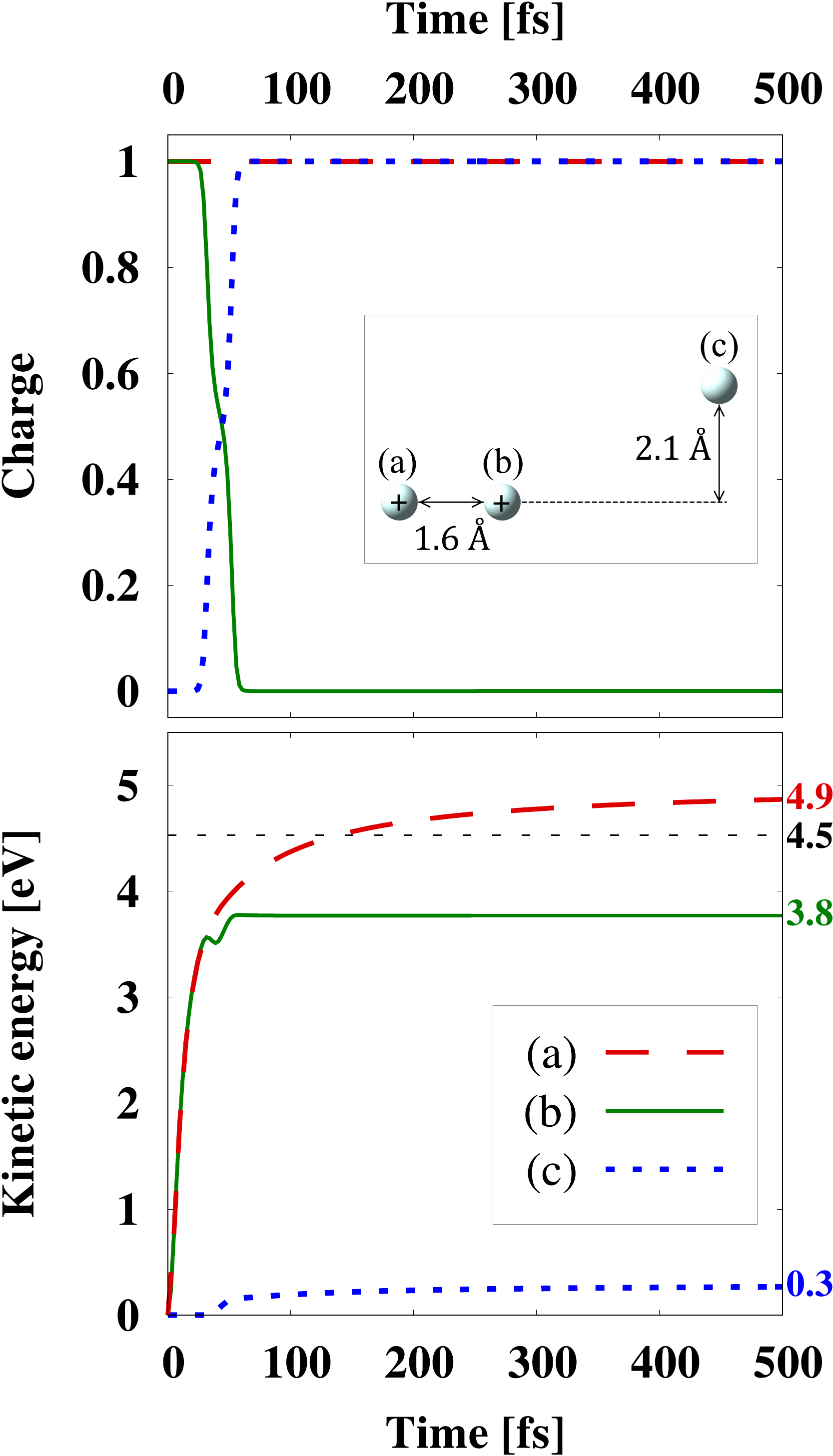}
	\centering
	\caption{\label{fig:traj} Upper panel: temporal evolution of the charges during a Coulomb explosion between the ions (a) and (b), with (b) transferring its charge to a neutral atom (c). Lower panel: kinetic energies of the three involved particles. The horizontal line at a kinetic energy of 4.5 eV corresponds to the asymptotic energy of the ions after a two-body Coulomb explosion.}
\end{figure}

We examined the trajectories of ions of such events using our theoretical model. One characteristic example of such trajectories is described here: initially, as ICD occurs, the two ions (labeled (a) and (b)), are generated at an interatomic distance of 1.6~\AA~(which is the inner turning point of the $\mathrm{He}_2^{+*}(2\mathrm{p})$ dimer potential). A third, neutral, atom, denoted as (c), is located at 5.0~\AA~from ion (b) and $2.1$~\AA~from the (a)-(b) internuclear axis. The upper panel of Fig. \ref{fig:traj} shows the temporal evolution of the charge of these particles. During the first thirty femtoseconds, the two ions repel each other with no influence from (c). Then, as (b) gets closer to (c), the positive charge is delocalized over the two atoms. At $t=44$~fs, which corresponds to the isosceles geometry with equal (ab) and (ac) distances, both atoms bear the same partial charge of $0.5$. From $t=44$~fs onwards, the charge of (b) decreases and becomes zero at around $t=60$~fs. The charge is thus transferred and the two particles (a) and (c) are ions from now on.  Additionally, the lower panel shows the change of the kinetic energy of these particles over time. It depicts, that the charge transfer takes place with almost no exchange of kinetic energy between (b) and (c) (less than 0.3~eV). After the charge transfer, the neutralized particle (b) keeps a nearly constant kinetic energy. On the contrary, now (a) and (c) repel each other due to their Coulomb repulsion. As (a) is much faster than (c) and the kinetic energy scales with the square of the velocity, (a) gains the majority of the available energy from this repulsion and reaches up to 5.0~eV kinetic energy during the whole process. Note that this is even more than what can be gathered from a two-body Coulomb explosion starting at 1.6~\AA. This extended energy range is well confirmed by the experiment and the theory as shown in  Fig. \ref{fig:map} where the non-diagonal part of the correlation spectra reaches ion kinetic energies up to 5.5~eV while the diagonal part does barely exceed 4.5~eV.

\begin{figure}
	\centering
	\includegraphics[scale=0.4,width=\linewidth]{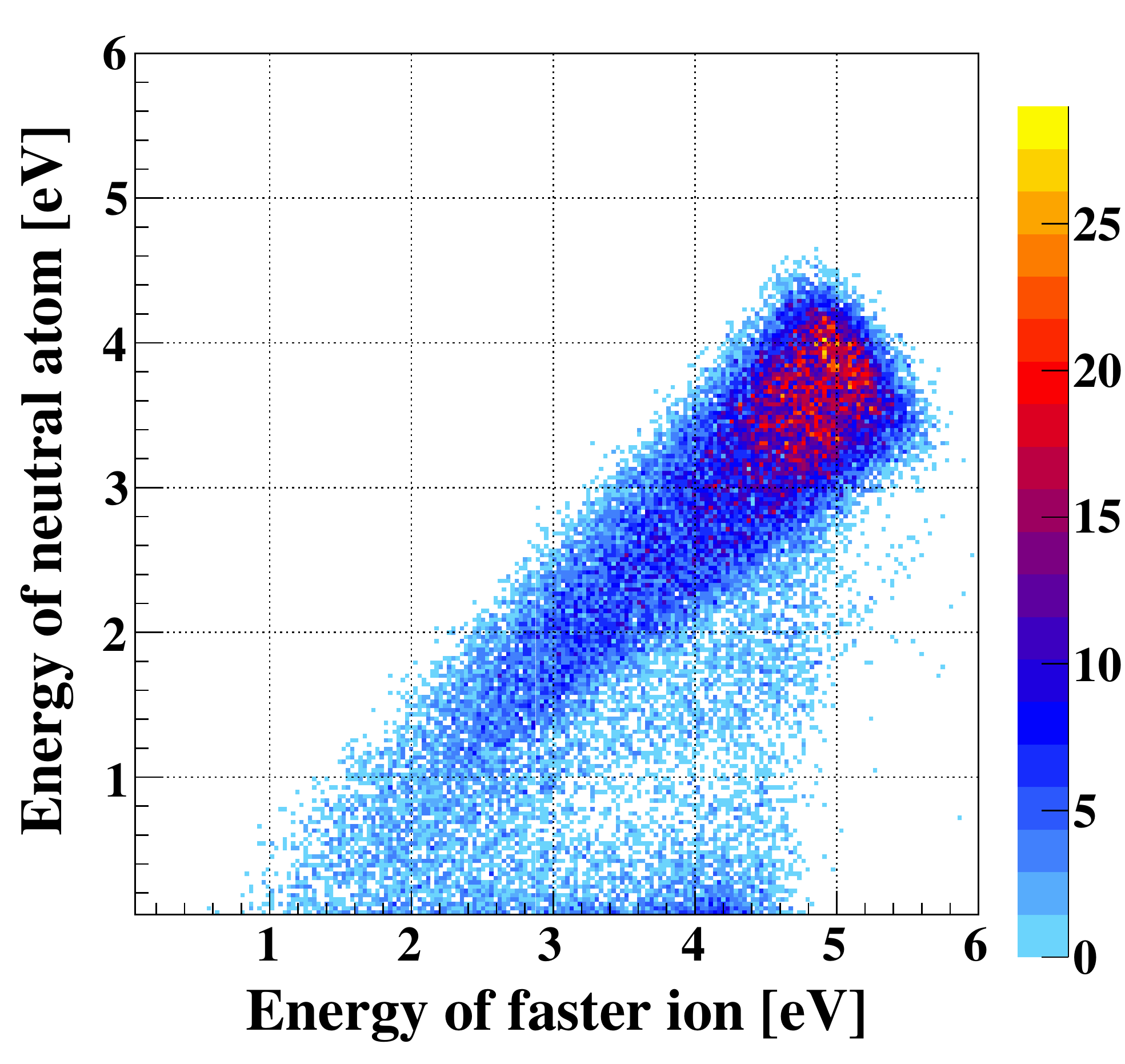}
	\caption{\label{fig:neutral} Computed energy distribution of the neutral He atom as function of the kinetic energy of the faster He$^+$ ion.}
\end{figure}

Finally, we show in Fig. \ref{fig:neutral} the computed~energy of the neutral atom as a function of that of the faster ion. A nearly diagonal feature is recovered since, as shown above, there is only a weak exchange of kinetic energy during the charge hopping process. After the frustrated Coulomb explosion, neutral helium atoms with kinetic energies up to 4.5 eV are formed.

To conclude, our theoretical and experimental study shows the occurrence of a charge hopping process during a Coulomb explosion of a small He cluster which leads to energetic neutral particles (with energies of several eV). Since the charge hopping probability increases with the cluster size, multiple charge transfers are thus possible for larger clusters and the neutral atoms might as a result gain the majority of the available kinetic energy. It is expected that this process is a very common route of fragmentation of loosely bound matter.

\section{Acknowledgement}
This project has received funding from the Research Executive Agency (REA) under the European Union's Horizon 2020 research and innovation programme Grant agreement No.\ 705515 and from Agence Nationale de la Recherche  through the program  ANR-16-CE29-0016-01. R.D., J.R., N.S. and T.J. acknowledge funding from the Deutsche Forschungsgemeinschaft (DFG) as part of the DFG research unit FOR1782/II.

\bibliography{helium}


\end{document}